\documentclass[11pt]{article}
\usepackage[margin=0.9in]{geometry}
\usepackage{amsmath,mathtools,amssymb,amsthm}
\usepackage{graphicx,bm,microtype}
\usepackage{xcolor,float,booktabs}
\usepackage[round]{natbib}
\usepackage[colorlinks=true,linkcolor=blue,citecolor=blue,urlcolor=blue,filecolor=blue]{hyperref}
\hypersetup{pdftitle={Scalable composite likelihood inference for directed probit networks},pdfauthor={Yibo Wang, Chenlei Leng, Guosheng Yin}}
\theoremstyle{plain}
\newtheorem{theorem}{Theorem}

\newtheorem{proposition}{Proposition}
\newtheorem{corollary}{Corollary}
\theoremstyle{definition}
\newtheorem{assumption}{Assumption}
\theoremstyle{remark}
\newtheorem{remark}{Remark}

\newcommand{\E}{\operatorname{E}}

\newcommand{\ind}{\mathbf1}
\newcommand{\R}{\mathbb R}
\newcommand{\argmax}{\operatorname*{arg\,max}}
\newcommand{\pto}{\xrightarrow{p}}
\newcommand{\dto}{\xrightarrow{d}}

\title{\bfseries Scalable composite likelihood inference for directed probit networks}
\author{Yibo Wang\thanks{Department of Statistics and Actuarial Science, The University of Hong Kong, Pokfulam Road, Hong Kong. Email: \href{mailto:yibo.wang@connect.hku.hk}{yibo.wang@connect.hku.hk}.}
\and Chenlei Leng\thanks{Department of Applied Mathematics, The Hong Kong Polytechnic University, Hung Hom, Kowloon, Hong Kong. Email: \href{mailto:chenlei.leng@polyu.edu.hk}{chenlei.leng@polyu.edu.hk}.}
\and Guosheng Yin\thanks{Department of Statistics and Actuarial Science, The University of Hong Kong, Pokfulam Road, Hong Kong. Email: \href{mailto:gyin@hku.hk}{gyin@hku.hk}.}}
\date{}
\begin{document}
\maketitle

\begin{abstract}
We develop scalable estimation and inference for directed probit networks with correlated sending and receiving effects and residual reciprocity. An incomplete composite likelihood in two stages combines exact bivariate margins of sampled dyads and triples, avoiding integration in high dimensions and cubic enumeration. For fixed parameter dimension, criterion evaluation is linear in the number of draws. Averaging a fixed number of fits from independent index samples reduces sampling variation and allows parallel computation. We establish consistency and joint asymptotic normality with different directional rates, allowing a singular node projection when draw counts lie between the node and dyad scales. An overlap covariance accounts for network and index variation and is consistent relative to each directional variance. Independent index samples stabilize covariance estimation without refitting. The resulting procedure provides standard errors, mapped joint regions and score tests for zero variance components. Simulations show improved estimation from averaging and substantially lower computation times than default Bayesian fitting. An application to a fly brain network with 21,738 neurons demonstrates estimation and inference at a scale relevant to modern network data.
\end{abstract}

\section{Introduction}
\label{sec:introduction}

Directed networks often exhibit unequal sending and receiving activity together with reciprocity, the tendency for two nodes to connect in both directions. After accounting for observed covariates, reciprocal association can arise both from nodes that tend to send and receive edges and from dependence specific to the two directions within a pair. Gaussian probit models with random effects distinguish these sources through correlated sending and receiving effects and reciprocal errors \citep{hoff2005bilinear,hoff2021additive}. We study estimation and inference for the regression coefficients, variances and covariance of the node effects, and residual reciprocity from one large network.

\citet{feng2025reciprocity} study reciprocity under different sparsity regimes and develop likelihood inference with node and dyad covariates. For $p_1$ models with fixed node effects \citep{holland1981exponential}, \citet{feng2026optimal} construct rate-optimal estimators and optimal tests for a common reciprocal effect using generalized cycle counts. These counts require $O(n^2d_{\max})$ operations, where $d_{\max}$ is the largest in-degree, out-degree or mutual degree. \citet{feng2025regression} allow reciprocity to depend on dyadic covariates and remove fixed node effects through a conditional likelihood based on tetrads. Enumerating and checking all tetrads requires $O(n^4)$ operations.

For the Gaussian model with random effects, each likelihood evaluation on a network with $n$ nodes involves a $2n$-dimensional integral. Crossed random effects also create dependence that cannot generally be partitioned into independent clusters. \citet{jiang2013subset} establishes consistency of maximum likelihood in generalized linear mixed models with crossed random effects through a subset argument. \citet{lyu2024increasing} develop asymptotic theory for maximum likelihood and restricted maximum likelihood in two-way crossed linear mixed models with increasing dimensions. These results provide foundations for inference with dependence from crossed effects.

Composite likelihood reduces the integration dimension by combining margins of lower dimension \citep{varin2011overview}. For probit models with crossed random effects by assuming independent row and column effects and independent latent errors, \citet{bellio2025crossed} construct hierarchical composite likelihoods involving integrals in one dimension, with computation linear in the number of observations per iteration. 

In directed networks, exact bivariate margins capture node heterogeneity and reciprocity, but enumerating all edge pairs that share a node requires $O(n^3)$ work. Randomly selecting likelihood components offers a way to reduce this cost. \citet{dillon2010stochastic} study stochastic composite likelihood and the trade-off between computation and statistical accuracy. Component selection also matters, as \citet{xu2024note} show that adding components or using margins of higher dimension need not improve efficiency.

The effect of sampling is closely related to incomplete $U$-statistics. \citet{chen2019randomized} establish Gaussian approximations with degenerate or nondegenerate kernels, and \citet{sturma2024testing} apply this approach to constraints under mixed degeneracy. \citet{shao2025reduction} study higher-order risk control and the statistical--computational trade-off under $U$-statistic reduction. In a network, sampled contributions share node effects and dyad errors, covariates vary across pairs, and staged estimation substitutes fitted parameters. Inference must account for these sources of dependence and uncertainty jointly.

We develop an incomplete composite likelihood in two stages using exact bivariate margins of uniformly sampled dyads and centred triples. Averaging a fixed number of fits reduces index variation and permits parallel computation. For fixed parameter dimension and number of fits, criterion evaluation is linear in the number of draws. We establish consistency and joint asymptotic normality with different directional rates. For comparable draw counts per fit of order $m$, the recommended range $n\ll m\ll n^2$ supports joint normality even when the leading node projection is singular. An overlap covariance accounts for network and index variation and is consistent relative to each directional variance; fresh index samples stabilize its estimation without refitting. The resulting inference includes standard errors, mapped joint regions and score tests for zero variance components. Simulations show gains in precision from averaging and substantial computational savings relative to default Bayesian fitting. An application to a fly brain network with 21,738 neurons demonstrates inference on a large network with correlated sending and receiving effects and residual reciprocity.

\section{Model and incomplete estimation}
\label{sec:estimation}
\subsection{Model and working parameters}
\label{sec:network-model}

For distinct nodes $i,j\in\{1,\ldots,n\}$, let $Y_{ij}$ indicate a directed edge and let $\bm x_{ij}\in\R^p$ be its covariate vector, with $p$ the number of covariates. We condition on $\bm X=\{\bm x_{ij}:i\ne j\}$ and consider
\begin{equation}
Z_{ij}=\bm x_{ij}^{\top}\bm\beta+a_i+b_j+\varepsilon_{ij},\qquad Y_{ij}=\ind\{Z_{ij}>0\},
\label{eq:model}
\end{equation}
where $a_i$ and $b_i$ describe node $i$'s tendencies to send and receive edges, respectively. With $\bm W_i=(a_i,b_i)^{\top}$, we assume
\[
\bm W_i\sim N_2\!\left(\bm0,\begin{pmatrix}\sigma_a^2&c\\c&\sigma_b^2\end{pmatrix}\right),\qquad (\varepsilon_{ij},\varepsilon_{ji})^{\top}\sim N_2\!\left(\bm0,\begin{pmatrix}1&\rho\\\rho&1\end{pmatrix}\right).
\]
Conditional on $\bm X$, the node vectors, $i=1,\ldots,n$, are independent, the error pairs indexed by $1\le i<j\le n$ are independent, and the two collections are independent of each other. The covariance $c$ describes association between a node's sending and receiving effects. When $c>0$, nodes with a greater residual tendency to send edges also tend to receive more edges. The correlation $\rho$ measures residual reciprocity conditional on covariates and node effects. A positive value $\rho>0$ indicates positive association between the two edge directions, while $\rho=0$ gives conditional independence. Only the signs of the latent variables are observed, so we fix the error variance to identify their scale. The parameter vector is $\bm{\lambda}=(\bm\beta^{\top},\rho,\sigma_a^2,\sigma_b^2,c)^{\top}\in\Lambda$, where $\Lambda$ consists of vectors with $\bm\beta\in\R^p$, $\sigma_a^2,\sigma_b^2\ge0$, $c^2\le\sigma_a^2\sigma_b^2$ and $|\rho|<1$.

Covariates describe the sender, receiver or ordered pair, and random effects represent remaining sending and receiving heterogeneity. The coefficients $\bm\beta$ act on the conditional latent scale; integrating out the effects gives the marginal regression scale below. Edges on disjoint node sets are independent. Shared nodes and reciprocal dyads determine the dependence captured by the four pair margins.

\begin{figure}[t]
\centering
\includegraphics[width=\textwidth]{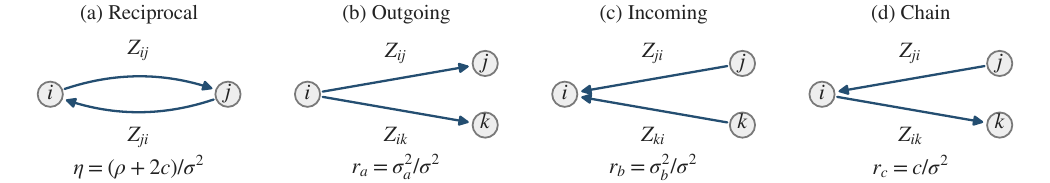}
\caption{Four base units with latent edge pairs and correlations, with $\sigma^2=1+\sigma_a^2+\sigma_b^2$. The other chain orientation interchanges $j$ and $k$.}
\label{fig:latent-pair-motifs}
\end{figure}

Conditional on $\bm X$, every latent edge variable has variance $\sigma^2=1+\sigma_a^2+\sigma_b^2$. Dividing by $\sigma$ preserves the binary outcomes and gives unit marginal variance. The standardized regression coefficients and marginal correlations of latent edge pairs are
\begin{equation}
\bm\gamma=\bm\beta/\sigma,\quad \eta=(\rho+2c)/\sigma^2,\quad
r_a=\sigma_a^2/\sigma^2,\quad r_b=\sigma_b^2/\sigma^2,\quad r_c=c/\sigma^2.
\label{eq:standard}
\end{equation}
The marginal edge probability is $\Pr(Y_{ij}=1\mid\bm X)=\Phi(\bm x_{ij}^{\top}\bm\gamma)$, where $\Phi$ is the standard normal distribution function. The correlations in \eqref{eq:standard} are also conditional on $\bm X$, with node effects integrated out. Specifically, $\eta$ is the correlation between $Z_{ij}$ and $Z_{ji}$ in a reciprocal pair. Their covariance is $\rho+2c$, with $\rho$ contributed by the reciprocal errors and $c$ by the sending and receiving effects of each of the two nodes. For distinct $i,j,k$, $r_a$ is the correlation between $Z_{ij}$ and $Z_{ik}$, which share sender $i$; $r_b$ is the correlation between $Z_{ji}$ and $Z_{ki}$, which share receiver $i$; and $r_c$ is the correlation between $Z_{ji}$ and $Z_{ik}$, which form a directed chain through $i$. Figure~\ref{fig:latent-pair-motifs} summarizes these four configurations, the nodes shared by each edge pair and the corresponding latent correlations.

Write $\bm r=(r_a,r_b,r_c)^{\top}$ and $\bm\theta=(\bm\gamma^{\top},\eta,\bm r^{\top})^{\top}$. The working parameter space $\Theta$ consists of these vectors with $\bm\gamma\in\R^p$, $r_a,r_b\ge0$, $r_a+r_b<1$, $r_c^2\le r_ar_b$ and $|\eta-2r_c|<1-r_a-r_b$. The inverse map $g:\Theta\to\Lambda$ is
\begin{equation}
\sigma^2=(1-r_a-r_b)^{-1},\quad \bm\beta=\sigma\bm\gamma,\quad
\sigma_a^2=\sigma^2r_a,\quad \sigma_b^2=\sigma^2r_b,\quad c=\sigma^2r_c,\quad \rho=\sigma^2(\eta-2r_c).
\label{eq:inverse}
\end{equation}

\subsection{Pair probabilities and sampled criteria}
\label{sec:likelihood-pairs}

For an interior parameter, let $f_W(\cdot;\bm{\lambda})$ be the density of the node effects. Conditional dyad independence gives the full likelihood
\begin{equation*}
\mathcal L(\bm{\lambda})=\int_{\R^{2n}}\prod_{1\le i<j\le n}\Pr(Y_{ij}=y_{ij},Y_{ji}=y_{ji}\mid\bm w_i,\bm w_j,\bm X;\bm{\lambda})\prod_{i=1}^n f_W(\bm w_i;\bm{\lambda})\,d\bm w_1\cdots d\bm w_n.
\end{equation*}
This $2n$-dimensional integral makes direct likelihood maximization computationally demanding. Integrating the effects for a pair of latent edges instead gives an exact bivariate normal margin, with correlation determined by shared node effects and dyad errors.

For binary outcomes $y_{ij}\in\{0,1\}$, let $q_{ij}=2y_{ij}-1$. Conditional on $\bm X$, the joint cell probabilities for the four edge pairs in Figure~\ref{fig:latent-pair-motifs} are
\[
\begin{aligned}
\text{Reciprocal }(Y_{ij},Y_{ji}):&\quad\ \ \ \ p^D_{ij}(\bm\gamma,\eta)
  =\Phi_2(q_{ij}\bm x_{ij}^{\top}\bm\gamma,q_{ji}\bm x_{ji}^{\top}\bm\gamma;q_{ij}q_{ji}\eta),\\
\text{Outgoing }(Y_{ij},Y_{ik}):&\quad p^a_{i;jk}(\bm\gamma,r_a)
  =\Phi_2(q_{ij}\bm x_{ij}^{\top}\bm\gamma,q_{ik}\bm x_{ik}^{\top}\bm\gamma;q_{ij}q_{ik}r_a),\\
\text{Incoming }(Y_{ji},Y_{ki}):&\quad p^b_{i;jk}(\bm\gamma,r_b)
  =\Phi_2(q_{ji}\bm x_{ji}^{\top}\bm\gamma,q_{ki}\bm x_{ki}^{\top}\bm\gamma;q_{ji}q_{ki}r_b),\\
\text{Chain }(Y_{ji},Y_{ik}):&\quad p^c_{i;jk}(\bm\gamma,r_c)
  =\Phi_2(q_{ji}\bm x_{ji}^{\top}\bm\gamma,q_{ik}\bm x_{ik}^{\top}\bm\gamma;q_{ji}q_{ik}r_c),
\end{aligned}
\]
where $\Phi_2(u,v;r)$ is the standard bivariate normal distribution function with correlation $r$. Each expression gives the probability of the displayed edge pair taking the specified binary values, with the outcome arguments suppressed. For example, $p^D_{ij}(\bm\gamma,\eta)=\Pr(Y_{ij}=y_{ij},Y_{ji}=y_{ji}\mid\bm X)$ at $(\bm\gamma,\eta)$. Varying these values gives the four cell probabilities. Parameter arguments are omitted when clear from the context. Write $\phi$ and $\phi_2$ for the corresponding univariate and bivariate densities.

Reciprocal dyads identify $(\bm\gamma,\eta)$ but do not separate the sources of reciprocal association. Outgoing, incoming and chain pairs identify $r_a,r_b,r_c$, respectively, and \eqref{eq:inverse} recovers the original parameters. We therefore first estimate $(\bm\gamma,\eta)$ from dyads, then estimate $\bm r$ from centred triples with the Stage I estimates fixed and the joint parameter restrictions enforced.

Let $\mathcal D=\{\{i,j\}:1\le i<j\le n\}$ and $\mathcal T=\{(i,\{j,k\}):i,j,k\text{ distinct},\ j<k\}$, with sizes $N_D=n(n-1)/2$ and $N_T=n(n-1)(n-2)/2$. Labels $D,T$ indicate dyad and triple quantities. Fix a positive integer $L$, independent of $n$. For each $\ell=1,\ldots,L$, draw $D_u^{(\ell)}\in\mathcal D$, $u=1,\ldots,m_D$, and $T_u^{(\ell)}\in\mathcal T$, $u=1,\ldots,m_T$, uniformly with replacement. Conditional on $\bm X$, all $2L$ samples are mutually independent and independent of all node effects and dyad errors. Every fit uses the same observed network. Empty dyads and zero outcomes within triples are retained.

For each pair of index samples, Stage I maximizes
\begin{equation*}
\widetilde Q_D^{(\ell)}(\bm\gamma,\eta)=m_D^{-1}\sum_{u=1}^{m_D}\log p^D_{D_u^{(\ell)}}(\bm\gamma,\eta),\qquad
(\widetilde{\bm\gamma}^{(\ell)},\widetilde\eta^{(\ell)})\in\argmax_{\bm\gamma\in\R^p,\ |\eta|<1}\widetilde Q_D^{(\ell)}(\bm\gamma,\eta).
\end{equation*}
For Stage II, set
\begin{equation}
\widetilde Q_T^{(\ell)}(\bm\gamma,\bm r)=m_T^{-1}\sum_{u=1}^{m_T}\left[\log p^a_{i;jk}+\log p^b_{i;jk}+\frac{1}{2}\{\log p^c_{i;jk}+\log p^c_{i;kj}\}\right]_{(i,\{j,k\})=T_u^{(\ell)}}.
\label{eq:incomplete}
\end{equation}
Within each centred triple, the chains $j\to i\to k$ and $k\to i\to j$ contribute their average, with weight $1/2$ each. This makes the contribution unchanged when $j$ and $k$ are interchanged and gives the combined chain term total weight one. For fixed $\bm\gamma$, the outgoing, incoming and averaged chain terms depend only on $r_a,r_b,r_c$, respectively. We use equal weights across these three terms; multiplying a term by a positive constant leaves its separate maximizer unchanged.

Holding $(\widetilde{\bm\gamma}^{(\ell)},\widetilde\eta^{(\ell)})$ fixed, we maximize over the feasible values of $\bm r$,
\begin{equation*}
\widetilde{\bm r}^{(\ell)}\in\argmax_{\bm r:((\widetilde{\bm\gamma}^{(\ell)})^{\top},\widetilde\eta^{(\ell)},\bm r^{\top})^{\top}\in\Theta}\widetilde Q_T^{(\ell)}(\widetilde{\bm\gamma}^{(\ell)},\bm r).
\end{equation*}
We can first perform three searches in one dimension. If their maximizers are jointly feasible, they maximize the Stage II criterion; otherwise, joint constrained optimization is required.

If a stage has more than one maximizer, we select one by a fixed deterministic rule, common to all fits. Writing $\widetilde{\bm\theta}^{(\ell)}=((\widetilde{\bm\gamma}^{(\ell)})^{\top},\widetilde\eta^{(\ell)},(\widetilde{\bm r}^{(\ell)})^{\top})^{\top}$, we average these working parameters and then transform them back to the original scale. When every fit exists in $\Theta$, the resulting estimator is
\begin{equation}
\bar{\bm\theta}=L^{-1}\sum_{\ell=1}^L\widetilde{\bm\theta}^{(\ell)},\qquad
\bar{\bm{\lambda}}=g(\bar{\bm\theta}).
\label{eq:averaged-fit}
\end{equation}
Taking $L=1$ gives a single sampled fit in two stages. For the theoretical statements, a component fit that does not exist is assigned a fixed interior value.

Conditional on the observed network, each sampled criterion has as its expectation the corresponding average over all dyads or centred triples. These complete criteria, defined in Section~\ref*{sec:full-reference} of the supplementary material, provide the theoretical reference for the sampled estimators.

Each fit keeps its sampled indices fixed during optimization and counts repeated draws with their multiplicities. For fixed $p$, evaluating all $L$ criteria or their derivatives requires $O\{L(m_D+m_T)\}$ local evaluations, or $O(n\log n)$ when both draw counts have that order and $L$ is fixed. The fits can run in parallel. Covariance construction uses score aggregation as described in Section~\ref{sec:averaged-overlap}; total time also depends on optimization steps and data access, with implementation details in Section~\ref*{sec:simulation-implementation}.

\begin{remark}[Two-stage and one-stage estimation]\label{rem:two-stage}
Estimation in two stages follows the identification structure. Reciprocal margins on two nodes identify $(\bm\gamma,\eta)$, and pair margins on three nodes identify the remaining dependence parameters. A similar separation of standardized regression and variance estimation underlies the all-row-column method of \citet{bellio2025crossed}. With the Stage I estimates fixed, Stage II permits three optimizations in one dimension when their maximizers are jointly feasible. The procedure avoids selecting a relative weight between stages and allows the tests for zero variance in Section~\ref{sec:zero-variance-test} to use the Stage I estimates and fresh triples alone. An alternative with one stage jointly solves
\[
\max_{\bm\theta\in\Theta}\left\{\widetilde Q_D^{(\ell)}(\bm\gamma,\eta)+\omega\widetilde Q_T^{(\ell)}(\bm\gamma,\bm r)\right\},\qquad \omega>0,
\]
where $\omega$ controls the relative contribution of triples. The paired comparison in Section~\ref*{sec:simulation-overall-summaries} of the supplementary material shows similar RMSEs for most parameters and similar marginal and joint coverage. Across matched settings, estimation in two stages reduces fitting time by 47--59\% and total fitting and inference time by 36--47\% relative to estimation in one stage. Estimation in one stage yields consistently lower RMSE only for the dyad covariate coefficient $\beta_4$, with small absolute gains. These results favour the simpler procedure in two stages for large network inference.
\end{remark}

\section{Large-sample properties}
\label{sec:large-sample}
\subsection{Identification and consistency}

We establish consistency and convergence rates as the network size and both draw counts per fit diverge. Joint normality follows when the leading variation also dominates the network and fitting remainders. We state this condition separately so that the dyad and triple draw counts may grow at different rates.

Unless stated otherwise, $\Pr$ and $\E$ are under the true law conditional on $\bm X$, including index randomness; further conditioning is explicit. The subscript 0 denotes truth, and $\Pr(\cdot\mid H_0)$ refers to a fixed parameter satisfying $H_0$. Norms are Euclidean for vectors and operator norms for matrices. Matrix inequalities use positive semidefinite order, and inverse square roots are symmetric.

\begin{assumption}[Network structure]\label{ass:sampling-structure}
The model and independence conditions in Section~\ref{sec:network-model} hold conditionally on $\bm X$, with fixed $p$.
\end{assumption}
\begin{assumption}[Interiority]\label{ass:parameter-space}
The true parameter is fixed, with $\bm\beta_0\in\R^p$, $0<\sigma_{a,0}^2,\sigma_{b,0}^2<\infty$, $c_0^2<\sigma_{a,0}^2\sigma_{b,0}^2$ and $|\rho_0|<1$; equivalently, $\bm\theta_0$ is interior to $\Theta$.
\end{assumption}
\begin{assumption}[Design]\label{ass:design}
The covariates are uniformly bounded and the design has full rank uniformly in $n$. For constants $K_x<\infty$ and $\kappa_x>0$, eventually
\[
\max_{i\ne j}\|\bm x_{ij}\|\le K_x,\qquad
\frac1{n(n-1)}\sum_{i\ne j}\bm x_{ij}\bm x_{ij}^{\top}\succeq\kappa_x\bm I_p.
\]
\end{assumption}
These conditions keep true pair probabilities positive and identify every regression direction. They describe fixed interior parameters and nonvanishing network density. For random designs independent of the latent effects and dyad errors, the design bounds must hold with probability tending to one, and additional covariance conditions must hold in probability.

\begin{proposition}[Identification]\label{prop:identification}
Under Assumptions~\ref{ass:sampling-structure}--\ref{ass:design}, for all sufficiently large $n$, equality of every reciprocal, outgoing, incoming and chain pair distribution under $\bm\theta\in\Theta$ and $\bm\theta_0$ implies $\bm\theta=\bm\theta_0$, and hence $\bm{\lambda}=\bm\lambda_0$.
\end{proposition}
\begin{assumption}[Index sampling]\label{ass:dyad-triple-subsampling}
The integer $L\ge1$ is fixed, and the numbers of draws per fit $m_D,m_T$ are positive integers specified in advance and both tend to infinity. Conditional on $\bm X$, each fitting index sample consists of independent uniform draws with replacement from its population. All $2L$ samples are mutually independent and independent of all node effects and dyad errors. All fits use the same network and the stagewise maximizers defined in Section~\ref{sec:likelihood-pairs}, with the same fixed rule for choosing among maximizers.
\end{assumption}
\begin{theorem}[Consistency and rate]\label{thm:incomplete}
Under Assumptions~\ref{ass:sampling-structure}--\ref{ass:dyad-triple-subsampling}, all $2L$ sampled stagewise maxima are simultaneously attained, and all component fits are interior, with probability tending to one. Moreover,
\[
\max_{1\le\ell\le L}\|\widetilde{\bm\theta}^{(\ell)}-\bm\theta_0\|
+\|\bar{\bm\theta}-\bm\theta_0\|+\|\bar{\bm\lambda}-\bm\lambda_0\|
=O_p\!\left\{\left(n^{-1}+m_D^{-1}+m_T^{-1}\right)^{1/2}\right\}.
\]
Whenever all component fits are feasible, their working mean belongs to $\Theta$.
\end{theorem}
Convexity of $\Theta$ preserves feasibility under averaging. Neither draw count needs to dominate $n$ for consistency. For $m_D\asymp m_T$ with $m=\max(m_D,m_T)$, the rate is $O_p\{(n^{-1}+m^{-1})^{1/2}\}$. Section~\ref*{sec:consistency-rate-proof} of the supplementary material gives the complete proof.

\subsection{Joint normal approximation}

For a dyad and a centred triple, respectively, define the following scores in $\R^{p+4}$
\[
\begin{aligned}
\bm v^D_{ij}(\bm\theta)
&=\left(\frac{\partial\log p^D_{ij}}{\partial\bm\gamma^{\top}},\frac{\partial\log p^D_{ij}}{\partial\eta},\bm0_3^{\top}\right)^{\top},\\
\bm v^T_{i;jk}(\bm\theta)
&=\left(\bm0_{p+1}^{\top},\frac{\partial}{\partial\bm r^{\top}}\left[\log p^a_{i;jk}+\log p^b_{i;jk}+\frac{1}{2}\{\log p^c_{i;jk}+\log p^c_{i;kj}\}\right]\right)^{\top}.
\end{aligned}
\]
For fit $\ell=1,\ldots,L$, the stacked estimating function is
\[
\widetilde{\bm\Psi}^{(\ell)}(\bm\theta)=m_D^{-1}\sum_{u=1}^{m_D}\bm v^D_{D_u^{(\ell)}}(\bm\theta)+m_T^{-1}\sum_{u=1}^{m_T}\bm v^T_{T_u^{(\ell)}}(\bm\theta).
\]
Write $\bm Y=\{Y_{ij}:i\ne j\}$. Uniform index sampling gives
\[
\E\{\widetilde{\bm\Psi}^{(\ell)}(\bm\theta)\mid\bm Y\}=\bm\Psi(\bm\theta)
=N_D^{-1}\sum_{e\in\mathcal D}\bm v_e^D(\bm\theta)+N_T^{-1}\sum_{t\in\mathcal T}\bm v_t^T(\bm\theta).
\]
Define
\[
\bm H=-\E \{\partial\bm\Psi(\bm\theta_0)/\partial\bm\theta^{\top}\},\qquad
\bar{\bm H}=-L^{-1}\sum_{\ell=1}^L\left.\frac{\partial\widetilde{\bm\Psi}^{(\ell)}(\bm\theta)}{\partial\bm\theta^{\top}}\right|_{\bm\theta=\bar{\bm\theta}}.
\]
Both sensitivities are block lower triangular. Stage I does not depend on $\bm r$, while the lower-left block propagates the Stage I uncertainty into Stage II. The fitted sensitivity uses all fitting samples at $\bar{\bm\theta}$. The estimator averages component fits; it need not solve the score equation obtained by combining their fitting samples.

For $i=1,\ldots,n$, define
\[
\bm\xi_i(\bm W_i)=n\E \{\bm\Psi(\bm\theta_0)\mid\bm W_i\},\qquad
\bm\Omega=n^{-1}\sum_{i=1}^n\E \{\bm\xi_i\bm\xi_i^{\top}\}.
\]
The covariance of the leading independent node sum is $\bm\Omega/n$. This variation is shared by all fits, but may vanish in some identifiable parameter directions. The coefficients for index sampling are
\[
\bm\Gamma_D=N_D^{-1}\sum_{e\in\mathcal D}\E \{\bm v^D_e(\bm v^D_e)^{\top}\},\qquad
\bm\Gamma_T=N_T^{-1}\sum_{t\in\mathcal T}\E \{\bm v^T_t(\bm v^T_t)^{\top}\},
\]
where the scores are evaluated at $\bm\theta_0$.
\begin{proposition}[Sampling information]\label{prop:sampling-information}
Under Assumptions~\ref{ass:sampling-structure}--\ref{ass:design}, the only nonzero blocks of $\bm\Gamma_D$ and $\bm\Gamma_T$ are their upper-left $(p+1)\times(p+1)$ and lower-right $3\times3$ blocks, respectively. The former equals the upper-left block of $\bm H$. There exist constants $0<c_\Gamma<C_\Gamma<\infty$ such that, eventually,
\[
c_\Gamma\bm I_{p+4}\preceq\bm\Gamma_D+\bm\Gamma_T\preceq C_\Gamma\bm I_{p+4}.
\]
The sensitivity is eventually nonsingular, with $\|\bm H\|+\|\bm H^{-1}\|$ uniformly bounded.
\end{proposition}
The two samples supply information on complementary score blocks even when $\bm\Omega$ is singular. Let $a_n=n^{-1}+m_D^{-1}+m_T^{-1}$. Each component fit admits a score expansion with remainder $O_p(a_n)$, and averaging a fixed number of these expansions preserves that rate.

The leading score and estimator covariance matrices are
\[
\bm K_L=\bm\Omega/n+\bm\Gamma_D/(Lm_D)+\bm\Gamma_T/(Lm_T),\qquad
\bm S_L=\bm H^{-1}\bm K_L\bm H^{-\top},\qquad
\kappa_L=\lambda_{\min}(\bm K_L).
\]
Only the index variation is divided by $L$; the observed network is shared. Matrix standardization permits different rates in different directions.
\begin{theorem}[Joint normality]\label{thm:incomplete-normality}
Under Assumptions~\ref{ass:sampling-structure}--\ref{ass:dyad-triple-subsampling}, $\bm K_L$ and $\bm S_L$ are eventually positive definite. The draw counts per fit may be unequal. If
\begin{equation}
a_n/\sqrt{\kappa_L}\longrightarrow0,
\label{eq:averaged-general-budget-condition}
\end{equation}
then
\[
\bm S_L^{-1/2}(\bar{\bm\theta}-\bm\theta_0)\dto N_{p+4}(\bm0,\bm I_{p+4}).
\]
For comparable draw counts $m_D\asymp m_T$, with $m=\max(m_D,m_T)$, condition~\eqref{eq:averaged-general-budget-condition} is equivalent to
\begin{equation}
n\lambda_{\min}(\bm\Omega)+n^2/(Lm)\longrightarrow\infty.
\label{eq:general-budget-condition}
\end{equation}
\end{theorem}
The theorem allows singular $\bm\Omega$ and requires no convergence of $\bm\Omega$, $\bm H$ or the sampling coefficients. Section~\ref*{sec:main-normality-proof} gives the proof using the joint normal approximation in Section~\ref*{sec:unified-inference-proof}, which combines the shared network term with the independent index errors, conditionally on the network.

\begin{corollary}[Recommended sampling range]\label{cor:recommended-budgets}
Under Assumptions~\ref{ass:sampling-structure}--\ref{ass:dyad-triple-subsampling}, if $m_D\asymp m_T$, $m/n\to\infty$ and $m/n^2\to0$, the normal limit in Theorem~\ref{thm:incomplete-normality} holds even when $\bm\Omega$ is singular. There exist constants $0<c_S<C_S<\infty$, possibly depending on fixed $L$, such that, eventually,
\[
c_S(Lm)^{-1}\bm I_{p+4}\preceq\bm S_L\preceq C_Sn^{-1}\bm I_{p+4}.
\]
\end{corollary}
We recommend this intermediate range. Its lower bound on the number of draws makes sampling variation negligible in directions with node variance of order $n^{-1}$; its upper bound keeps the $O_p(n^{-1})$ remainder below the smallest leading standard deviation. Choices include $n\log n$ and $n^{1+\delta}$ for fixed $0<\delta<1$, with fixed positive multipliers. For fixed $p$ and $L$, each evaluation of all fitting criteria or their derivatives requires $O(m)$ work, which is subquadratic in this range and $O(n\log n)$ when $m\asymp n\log n$.

More generally, all comparable diverging draw counts $m=o(n^2)$ satisfy~\eqref{eq:general-budget-condition}. For quadratic and larger comparable draw counts, the sufficient condition is equivalent to $n\lambda_{\min}(\bm\Omega)\to\infty$. Section~\ref*{sec:general-budget-results} gives the detailed rate classification and limit results beyond this condition.

Writing $\bm S_1$ for the leading covariance of a single fit, we have
\begin{equation}
\bm S_1-\bm S_L
=(1-L^{-1})\bm H^{-1}\left(\frac{\bm\Gamma_D}{m_D}+\frac{\bm\Gamma_T}{m_T}\right)\bm H^{-\top}\succeq0.
\label{eq:averaged-leading-gain}
\end{equation}
At fixed draw counts per fit, averaging reduces index variation. At fixed total counts $Lm_D,Lm_T$, it has the same leading covariance as one fit to the pooled occurrences, while allowing separate computation. These results concern fixed $L$; growing $L$ requires uniform control of failures and remainders.

The supplementary results also include complete criterion inference and potentially non-Gaussian limits at quadratic draw counts. Gaussian and chi-squared calibration here uses condition~\eqref{eq:averaged-general-budget-condition}.

\section{Covariance estimation and inference}
\label{sec:inference}
\subsection{Pooled and fresh averaged overlap covariances}
\label{sec:averaged-overlap}

All fits share the same network, so covariance estimation must retain dependence across fitting groups as well as between stages. At a fixed parameter, scores on disjoint node sets are independent, which allows us to collect the relevant products through aggregates over node subsets. For a dyad or triple $t$, let $V(t)$ be its node set. For nonempty $\mathcal A\subseteq\{1,\ldots,n\}$ with $|\mathcal A|\le3$, form, at a common argument,
\begin{equation}
\widetilde{\bm C}_{\mathcal A}^{(\ell)}(\bm\theta)=m_D^{-1}\sum_{u=1}^{m_D}\ind\{\mathcal A\subseteq V(D_u^{(\ell)})\}\bm v^D_{D_u^{(\ell)}}(\bm\theta)+m_T^{-1}\sum_{u=1}^{m_T}\ind\{\mathcal A\subseteq V(T_u^{(\ell)})\}\bm v^T_{T_u^{(\ell)}}(\bm\theta).
\label{eq:sampled-aggregates}
\end{equation}
With $\epsilon_{\mathcal A}=(-1)^{|\mathcal A|+1}$, set
\[
\bm C_{\mathcal A}^{\mathrm{pool}}(\bm\theta)=L^{-1}\sum_{\ell=1}^L\widetilde{\bm C}_{\mathcal A}^{(\ell)}(\bm\theta),\qquad
\bm J_L^{\mathrm{pool}}(\bm\theta)=\sum_{\mathcal A}\epsilon_{\mathcal A}\bm C_{\mathcal A}^{\mathrm{pool}}(\bm\theta)\bm C_{\mathcal A}^{\mathrm{pool}}(\bm\theta)^{\top},
\]
where sums over $\mathcal A$ range over all nonempty node subsets of size at most three, both here and below. The pooled aggregate combines all fitting samples, with denominators $Lm_D,Lm_T$. Its outer products retain contributions from different fitting groups and stages. The alternating signs implement inclusion--exclusion over shared node subsets, adapting the principle underlying multiway cluster covariance estimation \citep{cameron2011robust} to the local supports used here. Each product with intersecting node sets is counted exactly once, including self-products and repeated draws. Pair and triple terms remove the multiple counting in node aggregates. Each dyad updates three nonempty subsets and each triple seven, so storing only nonzero aggregates avoids enumerating all pairs of sampled contributions.

Our covariance construction builds on established approaches to inference with overlapping observations. \citet{aronow2015cluster} develop robust variance estimation for dyadic data that allows dependence between observations sharing an actor. \citet{cameron2011robust} develop multiway cluster covariance estimators using inclusion--exclusion to account for intersecting clustering dimensions. The present setting requires a related construction for local contributions supported on two or three nodes, where shared nodes may occupy different roles. Our overlap aggregation also accommodates covariance between stages and variation from index sampling, while avoiding explicit enumeration of all pairs of sampled configurations.

Evaluating both the overlap matrix and sensitivity at the parameter average gives the sandwich based on the fitting indices
\begin{equation}
\bm V_L^{\mathrm{pool}}=\bar{\bm H}^{-1}\bm J_L^{\mathrm{pool}}(\bar{\bm\theta})\bar{\bm H}^{-\top}.
\label{eq:incomplete-sandwich}
\end{equation}
To reduce index variation in covariance estimation, fix an integer $R\ge1$, independent of $n$, and draw $R$ additional groups of dyads and triples. We call these groups fresh because, conditional on the design, they are mutually independent and independent of all fitting indices and latent network variables. Each group contains $Lm_D$ dyad draws and $Lm_T$ triple draws. Use these denominators in \eqref{eq:sampled-aggregates} to form $\bm C_{\mathcal A}^{[r]}$, where $r=1,\ldots,R$ labels a fresh group, and write
\[
\bm J_L^{[r]}(\bm\theta)=\sum_{\mathcal A}\epsilon_{\mathcal A}\bm C_{\mathcal A}^{[r]}(\bm\theta)\bm C_{\mathcal A}^{[r]}(\bm\theta)^{\top},\qquad
\bar{\bm J}=R^{-1}\sum_{r=1}^R\bm J_L^{[r]}(\bar{\bm\theta}).
\]
The corresponding sandwich is $\bar{\bm V}=\bar{\bm H}^{-1}\bar{\bm J}\bar{\bm H}^{-\top}$, using the same fitting sensitivity. The integer $L$ determines the parameter average and its leading covariance, whereas $R$ controls the averaging of covariance estimates at that parameter. Conditional on the network and all fitting indices, averaging divides each entry's variance from fresh indices by $R$ and leaves its conditional mean unchanged. This mean depends on the draw counts $Lm_D,Lm_T$. Combining all $R$ fresh groups into one larger sample would therefore change the sampling covariance target.

\begin{theorem}[Overlap covariance consistency]\label{thm:scalable-sandwich}
Under the conditions of Theorem~\ref{thm:incomplete-normality}, let $\bm J_L$ be either $\bm J_L^{\mathrm{pool}}(\bar{\bm\theta})$ or $\bar{\bm J}$ for fixed $R$, and write $\bm V_L=\bar{\bm H}^{-1}\bm J_L\bar{\bm H}^{-\top}$. Then
\[
\|\bm K_L^{-1/2}\bm J_L\bm K_L^{-1/2}-\bm I_{p+4}\|\pto0,\qquad
\|\bm S_L^{-1/2}\bm V_L\bm S_L^{-1/2}-\bm I_{p+4}\|\pto0.
\]
The fitted sensitivity is nonsingular, and both the raw overlap matrix and sandwich are positive definite, with probability tending to one. For $H_0:\bm\theta=\bm\theta_0$ and fixed $\alpha\in(0,1)$,
\[
\Pr\{(\bar{\bm\theta}-\bm\theta_0)^{\top}(\bm V_L)^{-1}(\bar{\bm\theta}-\bm\theta_0)>\chi^2_{p+4,1-\alpha}\mid H_0\}\longrightarrow\alpha.
\]
\end{theorem}
Relative consistency controls covariance error on each directional scale. The proof in Section~\ref*{sec:main-covariance-proof} combines the pooled draw counts with the local rate of $\bar{\bm\theta}$ through the component score expansions.

Average raw overlap matrices before any correction for negative eigenvalues, which is asymptotically inactive. The draw counts already account for parameter averaging, so no further division by $L$, $R$ or $n$ is needed. Averaging component covariances retains index variation from a single fit; dividing by $L$ also incorrectly reduces the shared network variation. For $L=1$, either construction estimates the covariance of a single fit.

\subsection{Inference on the original scale}
\label{sec:main-original-inference}

With $\bm G_g=g'(\bm\theta_0)$ and $\bar{\bm G}_g=g'(\bar{\bm\theta})$, set
\begin{equation}
\bm S_{\lambda,L}=\bm G_g\bm S_L\bm G_g^{\top},\qquad
\bm V_{\lambda,L}=\bar{\bm G}_g\bm V_L\bar{\bm G}_g^{\top}.
\label{eq:averaged-original-covariance}
\end{equation}
\begin{corollary}[Delta-method and Wald inference]\label{cor:averaged-original-scale}
Under the conditions of Theorem~\ref{thm:incomplete-normality}, either covariance scheme satisfies
\[
\bm S_{\lambda,L}^{-1/2}(\bar{\bm{\lambda}}-\bm\lambda_0)\dto N_{p+4}(\bm0,\bm I_{p+4}),\qquad
\|\bm S_{\lambda,L}^{-1/2}\bm V_{\lambda,L}\bm S_{\lambda,L}^{-1/2}-\bm I_{p+4}\|\pto0.
\]
For a fixed $\bm F:\Lambda\to\R^h$, twice continuously differentiable near $\bm\lambda_0$ with a Jacobian of full row rank there, write $\bar{\bm G}_F=\bm F'(\bar{\bm{\lambda}})$. For $H_0:\bm F(\bm{\lambda})=\bm0$ and fixed $\alpha\in(0,1)$,
\begin{equation}
\Pr\bigl\{\bm F(\bar{\bm{\lambda}})^{\top}(\bar{\bm G}_F\bm V_{\lambda,L}\bar{\bm G}_F^{\top})^{-1}\bm F(\bar{\bm{\lambda}})>\chi^2_{h,1-\alpha}\mid H_0\bigr\}\longrightarrow\alpha.
\label{eq:averaged-restriction-wald}
\end{equation}
\end{corollary}
The map and its inverse are smooth near the interior truth. Condition~\eqref{eq:averaged-general-budget-condition} controls the Taylor remainder after matrix standardization, allowing different directional rates. Scalar intervals are $\bar\lambda_j\pm z_{1-\alpha/2}\{(\bm V_{\lambda,L})_{jj}\}^{1/2}$ for $j=1,\ldots,p+4$, where $z_u$ is the standard normal quantile. Boundary hypotheses are treated separately in Section~\ref{sec:zero-variance-test}.

For joint inference on all parameters, map the region on the working scale through $g$ to retain the nonlinear shape of its boundary.
\begin{corollary}[Mapped joint region]\label{cor:averaged-mapped-joint-region}
Under the conditions of Theorem~\ref{thm:incomplete-normality}, fix $\alpha\in(0,1)$. When all component fits exist in the interior of $\Theta$, $\bar{\bm H}$ is nonsingular and $\bm V_L$ is positive definite, define
\[
\mathcal C_\theta=\{\bm\theta\in\Theta:(\bm\theta-\bar{\bm\theta})^{\top}(\bm V_L)^{-1}(\bm\theta-\bar{\bm\theta})\le\chi^2_{p+4,1-\alpha}\},\qquad
\mathcal C_\lambda=g(\mathcal C_\theta),
\]
and set both regions to be empty otherwise. For every realization and every $\bm\lambda\in\Lambda$, $\bm\lambda\in\mathcal C_\lambda$ if and only if $g^{-1}(\bm\lambda)\in\mathcal C_\theta$. Moreover, $\Pr(\bm\lambda_0\in\mathcal C_\lambda)=\Pr(\bm\theta_0\in\mathcal C_\theta)\to1-\alpha$.
\end{corollary}
Section~\ref*{sec:transformation-inference-proof} of the supplementary material proves both corollaries.

Delta-method standard errors give compact marginal summaries and support smooth contrasts. Mapping retains the curved parameter restrictions, with membership checked by applying $g^{-1}$ and evaluating the quadratic form on the working scale. We therefore use delta intervals for marginal summaries and favour mapping for joint inference on all parameters.

\subsection{Testing a zero variance component}
\label{sec:zero-variance-test}

Zero variance components lie on the boundary of the parameter space and require a separate test. The null $H_{0a}:\sigma_a^2=0$ forces $c=0$ but allows receiver heterogeneity. Under this null, two outgoing edges from the same node are independent, motivating the correlation score at zero. Write $p_{ij}(\bm\gamma)=\Pr(Y_{ij}=1\mid\bm X)=\Phi(\bm x_{ij}^{\top}\bm\gamma)$ for the marginal success probability and set
\[
e_{ij}(\bm\gamma)=\frac{\phi(\bm x_{ij}^{\top}\bm\gamma)\{Y_{ij}-p_{ij}(\bm\gamma)\}}{p_{ij}(\bm\gamma)\{1-p_{ij}(\bm\gamma)\}},\qquad
h_{a,t}(\bm\gamma)=e_{ij}(\bm\gamma)e_{ik}(\bm\gamma),\quad t=(i,\{j,k\}).
\]
The test uses the $L$ Stage I fits from Section~\ref{sec:likelihood-pairs}, with $\bar{\bm\gamma}=L^{-1}\sum_{\ell=1}^L\widetilde{\bm\gamma}^{(\ell)}$. At every fixed finite parameter in the model's parameter space, including zero variance components, their maxima are simultaneously attained with probability tending to one and $\|\bar{\bm\gamma}-\bm\gamma_0\|=O_p\{(n^{-1}+m_D^{-1})^{1/2}\}$. The boundary argument in Section~\ref*{sec:zero-variance-proof} establishes this rate using Stage I alone.

Independently of all fitting indices and latent network variables conditional on the design, draw one fresh group of $Lm_T$ triples $T_u$, $u=1,\ldots,Lm_T$, uniformly with replacement. At the averaged Stage I estimate define
\begin{equation}
S_a=\frac1{Lm_T}\sum_{u=1}^{Lm_T} h_{a,T_u}(\bar{\bm\gamma}),\quad
\widetilde q_a=\frac1{Lm_T}\sum_{u=1}^{Lm_T}\{h_{a,T_u}(\bar{\bm\gamma})-S_a\}^2,\quad
Z_a=\frac{\sqrt{Lm_T}S_a}{\sqrt{\widetilde q_a}}.
\label{eq:zero-variance-score}
\end{equation}
For $H_{0b}:\sigma_b^2=0$, use $h_{b,t}=e_{ji}e_{ki}$ and define $Z_b$ from the same fresh triples. Neither the overlap sandwich nor $R$ enters these tests. Positive sender or receiver heterogeneity creates positive association, motivating tests in the upper tail.

\begin{proposition}[Zero-variance tests]\label{prop:zero-variance-score}
Under Assumptions~\ref{ass:sampling-structure}, \ref{ass:design} and the dyad sampling and fixed $L$ provisions of Assumption~\ref{ass:dyad-triple-subsampling}, use $m_D$ dyad draws for each Stage I fit and the independent fresh group of $Lm_T$ test triples defined above, with $m_D,m_T\to\infty$. At a fixed finite parameter in the model's parameter space, if
\begin{equation}
\sqrt{Lm_T}\bigl(n^{-1}+m_D^{-1}\bigr)\longrightarrow0,
\label{eq:zero-test-budget}
\end{equation}
then, for every $z\in\R$, $\Pr(Z_a\le z\mid H_{0a})\to\Phi(z)$ and $\Pr(Z_b\le z\mid H_{0b})\to\Phi(z)$. Against fixed positive sender or receiver variance, the corresponding test in the upper tail is consistent without condition~\eqref{eq:zero-test-budget}. Stage II is not needed under either null or alternative.
\end{proposition}
For fixed $\alpha\in(0,1)$, each test rejects above $z_{1-\alpha}$ with asymptotic size $\alpha$. The Bonferroni test satisfies $\limsup_{n\to\infty}\Pr\{\max(Z_a,Z_b)>z_{1-\alpha/2}\mid H_{0a}\cap H_{0b}\}\le\alpha$, without an independence requirement. A test does not reject if a required Stage I fit or its studentization is unavailable; these events have vanishing probability.

For fixed $L$, condition~\eqref{eq:zero-test-budget} is equivalent to $m_T=o(n^2)$ and $m_T=o(m_D^2)$. For comparable diverging draw counts it reduces to $m=o(n^2)$ and includes the recommended range. Under either null, the complete test score has zero projection on each node and order $O_p(n^{-1})$; the condition makes this term and the Stage I substitution error negligible relative to variation from fresh sampling. Section~\ref*{sec:zero-variance-proof} proves the result, including the Stage I rate at structural boundaries. The simulations assess these tests for $L=1$ and $L=5$.

\section{Simulations}
\label{sec:simulations}
\subsection{Design}

We examine point estimation, marginal and joint coverage, and computation under different forms of network dependence. We generate $\bm x_{ij}=(1,u_i,v_j,w_{ij})^{\top}$, where all $u_i$, $v_j$ and $w_{ij}$ are mutually independent and uniform on $[-3^{1/2},3^{1/2}]$, independently of the latent effects. Thus the design includes sender, receiver and dyad covariates. Covariates and networks are regenerated in each repetition. The baseline has $\bm\beta=(-0.3,0.6,-0.5,0.4)^{\top}$ and $(\rho,\sigma_a^2,\sigma_b^2,c)=(0.35,0.5,0.3,0.12)$.

Six further scenarios retain the baseline values except for the indicated changes. Weak reciprocal dependence uses $(\rho,c)=(0.10,0.03)$; strong and negative reciprocity use $\rho=0.65$ and $-0.45$, respectively. Offsetting dependence uses $c=-0.15$, so the node covariance opposes the positive reciprocal error correlation. Strong heterogeneity uses $(\sigma_a^2,\sigma_b^2,c)=(1.0,0.8,0.25)$, and lower density uses $\beta_1=-0.8$. The regression slopes are unchanged. Each scenario uses 1,000 independent repetitions at $n=200,600,1000,1400$, giving 28,000 networks.

We use $m_D=m_T=\lceil10n\log n\rceil$ or $\lceil5n^{3/2}\rceil$. Both lie in the recommended range of Corollary~\ref{cor:recommended-budgets}. For each network and choice of draw counts, we compare $L=1$ with $L=5$, with the single fit forming the first component of the average of five fits. The remaining component fits use independent index samples from the same network. We average working parameters before transformation and evaluate the sensitivity at this average using all fitting samples. Both procedures use $R=5$ fresh overlap groups, each with $Lm_D$ dyads and $Lm_T$ centred triples. The draw count per fit is held fixed, so the averaged procedure uses five times as many fitting draws.

Marginal intervals use delta-method standard errors and normal cutoffs. Joint inference concerns all eight parameters on the original scale and compares the delta and mapped regions at nominal 95\% coverage. Coverage includes all attempts, with unavailable intervals or regions counted as uncovered. The estimation and joint inference summaries weight the seven scenarios equally; the tests for zero variance and Bayesian comparison use the settings specified below. Computations use R on a Linux server with two Intel Xeon Gold 6248R processors at 3.00 GHz. Network repetitions run in parallel, with one numerical thread per worker. The component fits within each network are evaluated sequentially. Section~\ref*{sec:simulation-implementation} gives further details.

\subsection{Point estimation and marginal inference}

Table~\ref{tab:simulation-scalar-average} compares estimation and marginal inference at $n=200,1400$ for $L=1,5$, both with $R=5$. RMSE decreases with network size, and averaging lowers RMSE for every parameter and draw count shown. The largest relative gains are for the dyad coefficient and reciprocity.

\begin{table}[!htbp]
\centering
\small
\setlength{\tabcolsep}{4pt}
\caption{Estimation and marginal coverage, averaged over seven scenarios}
\begin{tabular}{lrrrrrrrrr}
\toprule
 & & \multicolumn{4}{c}{$n=200$} & \multicolumn{4}{c}{$n=1400$} \\
 & & \multicolumn{2}{c}{$\lceil10n\log n\rceil$} & \multicolumn{2}{c}{$\lceil5n^{3/2}\rceil$} & \multicolumn{2}{c}{$\lceil10n\log n\rceil$} & \multicolumn{2}{c}{$\lceil5n^{3/2}\rceil$} \\
Parameter & $L$ & RMSE & CP & RMSE & CP & RMSE & CP & RMSE & CP \\
\midrule
$\beta_1$ & 1 & 77.09 & 95.07 & 77.01 & 94.83 & 28.88 & 95.16 & 28.60 & 95.31 \\
 & 5 & 75.74 & 94.87 & 75.85 & 94.69 & 28.39 & 95.31 & 28.45 & 95.36 \\
$\beta_2$ & 1 & 57.68 & 94.60 & 56.99 & 94.54 & 20.83 & 95.41 & 20.22 & 95.47 \\
 & 5 & 54.89 & 94.30 & 54.67 & 94.37 & 20.04 & 95.41 & 19.93 & 95.51 \\
$\beta_3$ & 1 & 45.94 & 95.04 & 45.18 & 95.06 & 17.12 & 94.86 & 16.61 & 94.87 \\
 & 5 & 43.33 & 94.49 & 43.08 & 94.60 & 16.41 & 94.66 & 16.34 & 94.66 \\
$\beta_4$ & 1 & 18.76 & 95.21 & 16.77 & 95.39 & 5.36 & 95.01 & 3.44 & 95.01 \\
 & 5 & 12.09 & 93.96 & 11.59 & 94.36 & 2.70 & 95.30 & 2.01 & 95.16 \\
$\rho$ & 1 & 60.96 & 96.44 & 53.55 & 95.94 & 19.30 & 95.03 & 12.12 & 95.36 \\
 & 5 & 31.57 & 94.67 & 28.67 & 94.59 & 8.97 & 95.31 & 5.98 & 94.84 \\
$\sigma_a^2$ & 1 & 88.93 & 93.34 & 83.35 & 93.06 & 31.76 & 94.51 & 27.50 & 94.70 \\
 & 5 & 71.05 & 92.59 & 69.91 & 91.83 & 26.68 & 94.56 & 25.97 & 93.96 \\
$\sigma_b^2$ & 1 & 68.21 & 93.84 & 64.39 & 93.44 & 24.01 & 94.53 & 19.50 & 94.47 \\
 & 5 & 50.59 & 91.67 & 49.25 & 91.01 & 18.35 & 94.03 & 17.34 & 94.31 \\
$c$ & 1 & 46.92 & 94.04 & 45.33 & 93.57 & 17.01 & 94.51 & 15.26 & 94.60 \\
 & 5 & 39.96 & 92.73 & 39.77 & 92.60 & 14.84 & 94.71 & 14.51 & 94.46 \\
\bottomrule
\end{tabular}
\label{tab:simulation-scalar-average}
\par\smallskip
\begin{minipage}{\linewidth}
\footnotesize
RMSE is the square root of the mean squared error averaged equally over seven scenarios, multiplied by $10^3$; CP is the corresponding average coverage (\%) of nominal 95\% delta intervals. Each scenario uses 1,000 networks, shared by both values of $L$. The column headings give the number of draws per stage in each fit. Both procedures use $R=5$, with each fresh covariance group containing $L$ times the draw count per fit.
\end{minipage}
\end{table}

At $n=1400$, averaging lowers RMSE for the dyad coefficient from 5.36 to 2.70 with $\lceil10n\log n\rceil$ draws and from 3.44 to 2.01 with $\lceil5n^{3/2}\rceil$ draws, and lowers RMSE for reciprocity from 19.30 to 8.97 and from 12.12 to 5.98, respectively, on the table's $10^3$ scale. Gains are smaller for the intercept and coefficients of node covariates, whose uncertainty contains more network variation. At this size, marginal coverage averaged over scenarios ranges from 93.96\% to 95.51\% across parameters, draw counts and both values of $L$. Section~\ref*{sec:simulation-averaged-fit-details} reports results for each scenario at all four network sizes.

\subsection{Joint inference and computational cost}

\begin{table}[!htbp]
\centering
\small
\setlength{\tabcolsep}{4pt}
\caption{Average joint coverage (\%) and mean computation time (seconds)}
\begin{tabular}{rlrrrrrr}
\toprule
 & & \multicolumn{3}{c}{$L=1$} & \multicolumn{3}{c}{$L=5$} \\
$n$ & Draws & Delta & Mapped & Time & Delta & Mapped & Time \\
\midrule
200 & $\lceil10n\log n\rceil$ & 91.67 & 93.76 & 0.74 & 88.21 & 89.71 & 3.54 \\
200 & $\lceil5n^{3/2}\rceil$ & 91.33 & 92.57 & 0.95 & 87.61 & 89.39 & 4.68 \\
600 & $\lceil10n\log n\rceil$ & 94.53 & 94.87 & 2.94 & 93.19 & 93.67 & 14.58 \\
600 & $\lceil5n^{3/2}\rceil$ & 93.23 & 93.97 & 5.66 & 92.56 & 93.00 & 28.12 \\
1000 & $\lceil10n\log n\rceil$ & 94.64 & 94.87 & 5.51 & 94.06 & 94.36 & 27.75 \\
1000 & $\lceil5n^{3/2}\rceil$ & 94.37 & 94.73 & 12.63 & 93.74 & 94.14 & 63.26 \\
1400 & $\lceil10n\log n\rceil$ & 94.19 & 94.49 & 8.29 & 94.21 & 94.40 & 41.90 \\
1400 & $\lceil5n^{3/2}\rceil$ & 94.43 & 94.81 & 21.89 & 93.71 & 94.13 & 109.04 \\
\bottomrule
\end{tabular}
\label{tab:simulation-joint-time}
\par\smallskip
\begin{minipage}{\linewidth}
\footnotesize
Coverage concerns nominal 95\% regions for eight parameters, with equal weight for seven scenarios and 1,000 networks per scenario. Both values of $L$ use $R=5$. Delta uses the Wald region on the original scale; Mapped transforms the region on the working scale. Time is elapsed time for all fitting and covariance calculations within one network, excluding generation and storage. The $L$ fits within each network were evaluated sequentially.
\end{minipage}
\end{table}

Table~\ref{tab:simulation-joint-time} shows that mapped joint coverage moves closer to 95\% from $n=200$ to $n=1400$. At $n=1400$, the scenario averages are 94.49--94.81\% for $L=1$ and 94.13--94.40\% for $L=5$. Mapping improves coverage over the delta region in each displayed comparison. The improvement with network size is particularly marked for $L=5$, whose mapped coverage at $n=200$ is 89.39--89.71\%.

The recorded fitting and inference time for $L=5$ is about five times that for $L=1$. Parallel computation is available across fits and fresh covariance groups, as illustrated for covariance construction in Section~\ref{sec:real-data}. Sections~\ref*{sec:simulation-joint-comparison}--\ref*{sec:simulation-overall-summaries} report all four network sizes and QQ plots, together with a comparison against covariance based on the fitting indices that shows improved coverage from fresh covariance averaging without refitting.

\subsection{Zero-variance tests}

We use the baseline regression, reciprocity and covariate settings with $(\sigma_a^2,\sigma_b^2,c)=(0,0.3,0),\allowbreak(0.5,0,0),\allowbreak(0,0,0)$ to assess size. Each setting uses 1,000 networks at each of the four sizes, shared by $L=1,5$ and both draw counts. Each Stage I fit uses $m_D$ dyad draws, and the independent fresh test sample contains $Lm_T$ triples, with $m_D=m_T$ as specified in Section~\ref{sec:simulations}. Power is assessed using the baseline networks and their Stage I estimates from the estimation experiment.

\begin{table}[!htbp]
\centering
\small
\caption{Rejection rates (\%) of nominal 5\% tests for zero variance}
\begin{tabular}{rlrrrrrr}
\toprule
 & & & \multicolumn{2}{c}{One component zero} & \multicolumn{3}{c}{Both components zero} \\
$n$ & Draws & $L$ & $H_{0a}$ & $H_{0b}$ & $H_{0a}$ & $H_{0b}$ & Joint \\
\midrule
200 & $\lceil10n\log n\rceil$ & 1 & 4.7 & 4.6 & 4.2 & 5.0 & 4.4 \\
200 & $\lceil10n\log n\rceil$ & 5 & 4.7 & 3.3 & 4.4 & 6.0 & 6.2 \\
200 & $\lceil5n^{3/2}\rceil$ & 1 & 5.0 & 4.1 & 4.3 & 6.0 & 5.1 \\
200 & $\lceil5n^{3/2}\rceil$ & 5 & 2.8 & 3.3 & 4.9 & 5.7 & 5.3 \\
600 & $\lceil10n\log n\rceil$ & 1 & 5.3 & 4.4 & 5.4 & 5.2 & 5.2 \\
600 & $\lceil10n\log n\rceil$ & 5 & 3.9 & 4.3 & 5.3 & 4.8 & 5.0 \\
600 & $\lceil5n^{3/2}\rceil$ & 1 & 3.6 & 3.9 & 4.9 & 5.1 & 5.3 \\
600 & $\lceil5n^{3/2}\rceil$ & 5 & 4.6 & 2.6 & 5.2 & 5.4 & 6.4 \\
1000 & $\lceil10n\log n\rceil$ & 1 & 5.0 & 4.5 & 6.7 & 4.3 & 5.9 \\
1000 & $\lceil10n\log n\rceil$ & 5 & 4.4 & 4.8 & 4.9 & 5.8 & 5.1 \\
1000 & $\lceil5n^{3/2}\rceil$ & 1 & 4.4 & 4.0 & 5.5 & 4.6 & 4.3 \\
1000 & $\lceil5n^{3/2}\rceil$ & 5 & 4.4 & 4.6 & 5.6 & 4.9 & 4.9 \\
1400 & $\lceil10n\log n\rceil$ & 1 & 4.5 & 6.3 & 5.4 & 4.5 & 6.1 \\
1400 & $\lceil10n\log n\rceil$ & 5 & 4.8 & 3.9 & 5.4 & 4.0 & 4.5 \\
1400 & $\lceil5n^{3/2}\rceil$ & 1 & 4.5 & 4.1 & 4.8 & 4.9 & 5.1 \\
1400 & $\lceil5n^{3/2}\rceil$ & 5 & 3.8 & 4.5 & 5.9 & 6.3 & 5.7 \\
\bottomrule
\end{tabular}
\label{tab:variance-test-size}
\par\smallskip
\begin{minipage}{\linewidth}
\footnotesize
The Draws column gives $m_D=m_T$. Under one zero component, $H_{0a}$ uses $(\sigma_a^2,\sigma_b^2,c)=(0,0.3,0)$ and $H_{0b}$ uses $(0.5,0,0)$. Under two zero components, the individual tests use $z_{0.95}$ and the joint Bonferroni test uses $z_{0.975}$. Each entry uses 1,000 networks. Each Stage I fit uses $m_D$ dyads, and the fresh test group contains $Lm_T$ triples. The Monte Carlo standard error of a component rate at 5\% is 0.69 percentage points.
\end{minipage}
\end{table}

Table~\ref{tab:variance-test-size} reports rejection rates under the three null settings. Rates are generally close to 5\% at larger network sizes for both values of $L$. The power experiment in Section~\ref*{sec:variance-test-simulations} gives rejection in every repetition for the component and joint tests at the positive baseline variances. These tests require only Stage I estimates and fresh triples.

\subsection{Comparison with Markov chain Monte Carlo}
\label{sec:mcmc-comparison}

We fit the same 1,000 baseline networks at each of $n=200,600$ with the additive probit model with rank zero in \texttt{amen}, using default priors and sampling settings \citep{hoff2021additive}. Estimates are posterior means, and marginal intervals are equal-tail 95\% credible intervals based on 400 saved draws. The joint region maps a ellipsoid on the working scale calibrated by the empirical 95th percentile of posterior Mahalanobis distances. The incomplete comparisons use $L=1$ and $L=5$, both with $R=5$. Section~\ref*{sec:mcmc-details} gives the sampling settings, diagnostics and comparisons for each parameter.

\begin{table}[!htbp]
\centering
\small
\setlength{\tabcolsep}{3pt}
\caption{Baseline overall RMSE, coverage and computation on 1,000 common networks at each size}
\begin{tabular}{lrrrrrrrr}
\toprule
 & \multicolumn{4}{c}{$n=200$} & \multicolumn{4}{c}{$n=600$} \\
Method or draws & RMSE & Marginal & Joint & Time & RMSE & Marginal & Joint & Time \\
\midrule
MCMC & 41.88 & 93.5--94.8 & 92.0 & 488.22 & 23.35 & 94.0--95.2 & 92.9 & 5304.97 \\
$\lceil10n\log n\rceil$, $L=1$ & 53.56 & 93.0--96.3 & 93.9 & 0.76 & 29.87 & 92.8--95.4 & 94.4 & 2.99 \\
$\lceil5n^{3/2}\rceil$, $L=1$ & 51.50 & 93.2--96.3 & 92.2 & 0.97 & 27.07 & 93.9--95.6 & 94.3 & 5.75 \\
$\lceil10n\log n\rceil$, $L=5$ & 45.18 & 91.5--95.9 & 89.4 & 3.65 & 25.44 & 93.0--95.6 & 93.1 & 14.86 \\
$\lceil5n^{3/2}\rceil$, $L=5$ & 44.78 & 90.0--95.5 & 90.0 & 4.77 & 24.77 & 93.4--96.4 & 93.0 & 28.68 \\
\bottomrule
\end{tabular}
\label{tab:mcmc-joint-time}
\par\smallskip
\begin{minipage}{\linewidth}
\footnotesize
RMSE is the square root of the unweighted mean MSE over all eight parameters on the original scale, multiplied by $10^3$. Marginal is the coverage range (\%) over eight nominal 95\% intervals; Joint is coverage (\%) of the mapped region for eight parameters; Time is mean fitting and inference time in seconds. Draw counts are per stage in each incomplete fit. All incomplete estimates use $R=5$; the five component fits for $L=5$ are sequential. Bayesian regions use posterior cutoffs. Times describe the implemented procedures under their recorded computing loads.
\end{minipage}
\end{table}

Table~\ref{tab:mcmc-joint-time} shows substantial computational savings on the baseline networks. At $n=600$, incomplete fitting and inference take 2.99--5.75 seconds for $L=1$ and 14.86--28.68 seconds for $L=5$, compared with 88.42 minutes for default Bayesian fitting. Averaging brings the overall RMSE closer to MCMC at both network sizes. At this size, incomplete mapped coverage is 93.0--94.4\%, compared with 92.9\% for MCMC, with marginal coverage close to 95\% across methods. At $n=200$, the lower joint coverage for $L=5$ follows the pattern in Table~\ref{tab:simulation-joint-time}.

\section{Application to the Hemibrain network}
\label{sec:real-data}

We use a large neuronal network to examine how spatial position and morphology relate to directed connectivity, and to distinguish these anatomical associations from residual reciprocity and node heterogeneity.

We analyse the Hemibrain v1.2 reconstruction of the adult fruit fly central brain \citep{scheffer2020connectome,janelia2020hemibrain}. The reconstruction provides directed synaptic connections and neuronal skeletons within an observed brain volume. We use the released adjacency table for 21,739 traced neurons that are not cropped and 3,550,403 directed connections. An edge from neuron $i$ to neuron $j$ indicates at least one recorded synapse with $i$ presynaptic and $j$ postsynaptic. We exclude connections from a neuron to itself and match skeletons by neuron identifier. One neuron lacks a skeleton and is removed before fitting. The resulting network contains 21,738 neurons and 3,550,061 directed edges, with density 0.7513\%. Unrecorded connections within this retained population are coded as zero.

The covariates describe spatial separation and neuronal size. Let $L_i$ be the total cable length of neuron $i$, summed over skeleton segments, and let $d_{ij}$ be the Euclidean distance between skeleton centroids weighted by cable length. The indexed length $L_i$ is distinct from the number of fits $L$. Set $L_0$ to the median cable length and $d_0$ to the median distance among 100,000 uniformly sampled directed node pairs, using geometry only. We use
\begin{equation}
\bm x_{ij}=\bigl(1,d_{ij}/(d_{ij}+d_0),L_i/(L_i+L_0),L_j/(L_j+L_0)\bigr)^{\top}.
\label{eq:hemibrain-design}
\end{equation}
These bounded transformations are invariant to the coordinate unit, and both scaling constants are fixed before fitting. Separate sender and receiver lengths allow different associations in the two roles. All predictors are constructed from the matched skeletons.

\begin{figure}[!htbp]
\centering
\includegraphics[width=\linewidth]{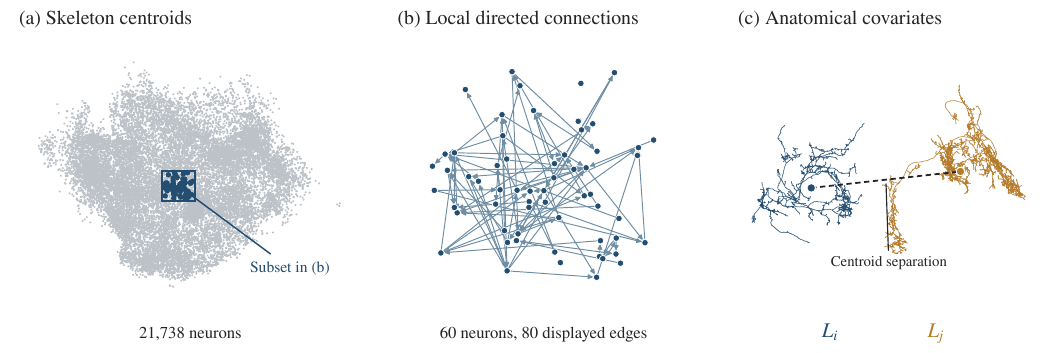}
\caption{The observed Hemibrain network and anatomical covariates. (a) Skeleton centroids of all 21,738 retained neurons. (b) A subset of 60 neurons selected by proximity to the coordinatewise median centroid; 80 of its 386 observed directed edges are selected uniformly for display. Arrowheads point from presynaptic to postsynaptic neurons. (c) Two example skeletons, with dots marking centroids weighted by cable length and a dashed segment showing their projected separation; $L_i,L_j$ denote total cable lengths. All panels use the first two principal components of the three-dimensional centroid coordinates, at different magnifications. The model uses the original three-dimensional distance $d_{ij}$.}
\label{fig:hemibrain-network}
\end{figure}

Figure~\ref{fig:hemibrain-network} shows the spatial distribution of skeleton centroids, directed connections in a local subset and two example skeletons used to illustrate the anatomical covariates.

We obtain $L=10$ fits from independent index samples on the same network and covariates, average the working parameters and then transform to the original scale. The primary numbers of draws per fit are $m_D=m_T=\lceil10n^{3/2}\rceil=32,050,100$. Table~\ref{tab:hemibrain-parameters} also reports averages of ten fits using $\lceil50n\log n\rceil$. We evaluate the pooled sensitivity at each average and use $R=5$ fresh overlap groups, each containing ten times as many dyads and triples as a component fit, as in Section~\ref{sec:averaged-overlap}.

\begin{table}[!htbp]
\centering
\small
\setlength{\tabcolsep}{3pt}
\caption{Hemibrain estimates from ten averaged fits and pointwise 95\% intervals}
\begin{tabular}{lrcrc}
\toprule
 & \multicolumn{2}{c}{$\lceil50n\log n\rceil$} & \multicolumn{2}{c}{$\lceil10n^{3/2}\rceil$} \\
Parameter & Estimate & Confidence interval & Estimate & Confidence interval \\
\midrule
Intercept & -2.59 & $[-2.63,-2.54]$ & -2.58 & $[-2.62,-2.54]$ \\
Distance & -4.99 & $[-5.05,-4.93]$ & -4.99 & $[-5.05,-4.93]$ \\
Sender cable length & 1.58 & $[1.52,1.64]$ & 1.58 & $[1.52,1.63]$ \\
Receiver cable length & 1.91 & $[1.88,1.95]$ & 1.91 & $[1.87,1.94]$ \\
Reciprocity ($\rho$) & 0.73 & $[0.72,0.74]$ & 0.72 & $[0.72,0.73]$ \\
Sender variance ($\sigma_a^2$) & 0.09 & $[0.08,0.10]$ & 0.09 & $[0.08,0.10]$ \\
Receiver variance ($\sigma_b^2$) & 0.04 & $[0.04,0.05]$ & 0.04 & $[0.04,0.05]$ \\
Node covariance ($c$) & 0.03 & $[0.03,0.04]$ & 0.03 & $[0.03,0.04]$ \\
\bottomrule
\end{tabular}
\label{tab:hemibrain-parameters}
\par\smallskip
\begin{minipage}{\linewidth}
\footnotesize
Each estimate averages ten fits on the working scale before transformation to the original scale. The numbers of draws per fit are 10,854,672 and 32,050,100 per stage. Intervals use delta-method standard errors from the pooled sensitivity and five fresh raw overlap matrices, each evaluated at the averaged parameter with ten times as many dyads and triples as a component fit.
\end{minipage}
\end{table}

Spatial separation has a strong negative association with recorded connectivity. At the same sender and receiver cable lengths, the model favours connections between neurons with closer skeleton centroids. Both length coefficients are positive. Thus neurons with more cable have a greater fitted tendency to send and receive connections, after accounting for spatial separation and their partners' cable lengths. These signs and magnitudes are stable under both choices of draw counts.

Marginal probabilities from the averaged working regression coefficients express these associations on the connection probability scale. When both cable lengths equal $L_0$, the fitted probabilities at distances $d_0/2,d_0,2d_0$ are 0.9348\%, 0.0866\%, 0.0045\%, respectively. At distance $d_0$, doubling the sender or receiver reference length gives fitted probabilities of 0.1951\% and 0.2301\%, respectively, with the partner length fixed. These comparisons based on the model show the magnitude of the associations at common covariate profiles.

Substantial reciprocity remains after adjustment for anatomy and node effects. With $\lceil10n^{3/2}\rceil$ draws per stage, the error correlation is 0.72, measuring residual dependence between the two directions within a neuron pair. The sender and receiver variances are 0.09 and 0.04, respectively, indicating greater residual heterogeneity in sending activity. The positive node covariance, 0.03, corresponds to a correlation between node effects of 0.54. Thus neurons with a greater residual tendency to send connections also tend to receive more connections. The analysis thus distinguishes dependence across a neuron's partners from reciprocity within an individual pair.

For each draw count, the tests for zero variance in Section~\ref{sec:zero-variance-test} use the regression average $\bar{\bm\gamma}$ from the corresponding ten Stage I fits and one independent group of $Lm_T=10m_T$ fresh triples. The sender and receiver statistics are $(Z_a,Z_b)=(33.48,18.04)$ for $m_D=m_T=\lceil50n\log n\rceil$ and $(59.65,29.57)$ for $m_D=m_T=\lceil10n^{3/2}\rceil$. Within each setting, the two tests use the same fresh triples. Both null hypotheses are rejected with Bonferroni-adjusted $p<0.001$ under either draw count, supporting residual heterogeneity in both sending and receiving activity after adjustment for anatomical covariates.

Both draw counts support the same substantive conclusions. The larger count gives narrower intervals for dependence parameters, while $50n\log n$ retains the interpretation at lower cost. Section~\ref*{sec:hemibrain-budget-sensitivity} assesses sensitivity of the averages of ten fits to the number of draws and to individual fitting index samples.

The complete population of centred triples has approximately $5.14\times10^{12}$ elements. With $50n\log n$ and $10n^{3/2}$ draws, mean fitting times for the two stages per component are 11.76 and 35.20 minutes, respectively. Running the ten fits concurrently takes 14.44 and 41.61 minutes; subsequent parallel covariance construction and interval calculation take 41.67 and 65.78 minutes. These elapsed times include all solver evaluations and exclude preprocessing. Section~\ref*{sec:hemibrain-averaged-inference} gives implementation and timing details.

\section{Discussion}
\label{sec:discussion}

Incomplete composite likelihood makes estimation and inference feasible for directed probit networks while retaining distinct parameters for node heterogeneity and residual reciprocity. Exact bivariate margins avoid integration in high dimensions, and sampling replaces cubic enumeration by a chosen number of local evaluations. For comparable draw counts per fit of order $m$, the recommended range $n\ll m\ll n^2$ combines this computational reduction with joint normal inference even when the leading node projection is singular. Relative consistency of the overlap covariance supports inference across different directional scales, and the score tests address zero variance components without optimization in the second stage.

Parameter averaging and covariance averaging control different sources of sampling variation. At fixed draw counts per fit, $L$ reduces leading index variation while retaining the shared network variation; $R$ stabilizes covariance estimation without refitting. Separate fits and fresh covariance groups can run in parallel, allowing computation to be allocated between estimation precision and covariance stability.

The simulations demonstrate substantial reductions in computation relative to the default Bayesian implementation, with single fits providing rapid inference and averaged fits improving precision, especially for reciprocity and the dyad covariate. Both achieve coverage close to nominal levels for larger networks. In the Hemibrain application, the method provides inference for 21,738 neurons and separates associations with anatomical distance and cable length from remaining reciprocal and node dependence. Agreement across draw counts supports the stability of the substantive conclusions.

Stratification based on covariates and unequal allocation between stages could further improve computational efficiency, with suitable weights and covariance adjustments. The present theory treats fixed numbers of fits, covariates of fixed dimension and fixed model parameters. Growing numbers of fits, vanishing edge probabilities and multiplicative latent effects are useful directions for further work. Incorporating synapse counts or partially observed connections would also broaden the scope of the network analysis.

The present theory assumes nonvanishing edge probabilities.
A sparse extension can be formulated by allowing the intercept
to diverge to $-\infty$ with the network size, while retaining
the exact bivariate probit margins. Such an extension requires
new information and sampling conditions because reciprocal pairs,
wedges and chains may occur at different rates, and uniform
sampling can yield predominantly contributions in which all outcomes are zero.
Sampling stratified by outcome with appropriate weighting may
reduce computational cost, but cannot compensate for insufficient
information in the complete network. Establishing limits in sparse regimes for the staged estimator and overlap covariance is therefore
a separate problem.
\section*{Declaration of the use of generative AI and AI-assisted technologies}
OpenAI Codex assisted with writing, simulation code and result analysis. Rethlas \citep{ju2026automated} with OpenAI GPT-6 Astra assisted with proof development. The author has independently verified the content and takes full responsibility for the manuscript.

\section*{Supplementary material}
Supplementary material contains proofs, further results for complete estimation and general numbers of draws, and implementation details. It also includes detailed simulation results and Bayesian comparisons, Hemibrain data processing and diagnostics, and an analysis of a Twitter following network.

\bibliographystyle{apalike}
\bibliography{paperbib}

@article{jiang2013subset,
  author  = {Jiang, Jiming},
  title   = {The subset argument and consistency of {MLE} in {GLMM}:
             Answer to an open problem and beyond},
  journal = {The Annals of Statistics},
  year    = {2013},
  volume  = {41},
  number  = {1},
  pages   = {177--195},
  url     = {https://arxiv.org/abs/1303.2874}
}

@article{lyu2024increasing,
  author  = {Lyu, Ziyang and Sisson, S. A. and Welsh, A. H.},
  title   = {Increasing dimension asymptotics for two-way crossed
             mixed effect models},
  journal = {The Annals of Statistics},
  year    = {2024},
  volume  = {52},
  number  = {6},
  pages   = {2956--2978},
  url     = {https://arxiv.org/abs/2401.06446}
}

@article{xu2024note,
  author  = {Xu, Libai and Reid, Nancy and Xu, Ximing},
  title   = {A note on information bias and efficiency of
             composite likelihood},
  journal = {Statistica Sinica},
  year    = {2024},
  volume  = {34},
  number  = {1},
  pages   = {523--526},
  doi     = {10.5705/ss.202022.0167},
  url     = {https://doi.org/10.5705/ss.202022.0167}
}

@article{dillon2010stochastic,
  author  = {Dillon, Joshua V. and Lebanon, Guy},
  title   = {Stochastic composite likelihood},
  journal = {Journal of Machine Learning Research},
  year    = {2010},
  volume  = {11},
  pages   = {2597--2633},
  url     = {https://www.jmlr.org/papers/volume11/dillon10a/dillon10a.pdf}
}

@article{aronow2015cluster,
  author  = {Aronow, Peter M. and Samii, Cyrus and
             Assenova, Valentina A.},
  title   = {Cluster-robust variance estimation for dyadic data},
  journal = {Political Analysis},
  year    = {2015},
  volume  = {23},
  number  = {4},
  pages   = {564--577},
  doi     = {10.1093/pan/mpv018},
  url     = {https://doi.org/10.1093/pan/mpv018}
}

@article{cameron2011robust,
  author  = {Cameron, A. Colin and Gelbach, Jonah B. and
             Miller, Douglas L.},
  title   = {Robust inference with multiway clustering},
  journal = {Journal of Business \& Economic Statistics},
  year    = {2011},
  volume  = {29},
  number  = {2},
  pages   = {238--249},
  doi     = {10.1198/jbes.2010.07136},
  url     = {https://doi.org/10.1198/jbes.2010.07136}
}

@article{hoff2005bilinear,
  author = {Hoff, Peter D.},
  title = {Bilinear Mixed-Effects Models for Dyadic Data},
  journal = {J. Am. Statist. Assoc.},
  year = {2005},
  volume = {100},
  number = {469},
  pages = {286--295},
  doi = {10.1198/016214504000001015}
}

@article{hoff2021additive,
  author = {Hoff, Peter D.},
  title = {Additive and Multiplicative Effects Network Models},
  journal = {Statist. Sci.},
  year = {2021},
  volume = {36},
  number = {1},
  pages = {34--50},
  doi = {10.1214/19-STS757}
}

@article{varin2011overview,
  author = {Varin, Cristiano and Reid, Nancy and Firth, David},
  title = {An overview of composite likelihood methods},
  journal = {Statist. Sinica}, year = {2011}, volume = {21}, number = {1}, pages = {5--42}
}

@article{chen2019randomized,
  author = {Chen, Xiaohui and Kato, Kengo},
  title = {Randomized incomplete {$U$}-statistics in high dimensions},
  journal = {Ann. Statist.}, year = {2019}, volume = {47}, number = {6}, pages = {3127--3156},
  doi = {10.1214/18-AOS1773}
}

@article{feng2025reciprocity,
  author = {Feng, Rui and Leng, Chenlei},
  title = {Modelling directed networks with reciprocity},
  journal = {Biometrika}, year = {2025}, volume = {112}, number = {2},
  pages = {asaf035}, doi = {10.1093/biomet/asaf035}
}

@article{bellio2025crossed,
  author = {Bellio, Ruggero and Ghosh, Swarnadip and Owen, Art B. and Varin, Cristiano},
  title = {Consistent and scalable composite likelihood estimation of probit models with crossed random effects},
  journal = {Biometrika},
  year = {2025}, volume = {112}, number = {3}, pages = {asaf037},
  doi = {10.1093/biomet/asaf037}
}

@article{scheffer2020connectome,
  author = {Scheffer, Louis K. and Xu, C. Shan and Januszewski, Michal and others},
  title = {A connectome and analysis of the adult {Drosophila} central brain},
  journal = {eLife}, year = {2020}, volume = {9}, pages = {e57443},
  doi = {10.7554/eLife.57443}
}

@misc{janelia2020hemibrain,
  author = {Katz, Bill},
  title = {{FlyEM} {Hemibrain} v1.2 release},
  year = {2020},
  howpublished = {DVID blog, Janelia Research Campus, \url{https://dvid.io/blog/release-v1.2/}}
}

@article{sturma2024testing,
  author = {Sturma, Nils and Drton, Mathias and Leung, Dennis},
  title = {Testing many constraints in possibly irregular models using incomplete {$U$}-statistics},
  journal = {J. R. Statist. Soc. {\rm B}},
  year = {2024}, volume = {86}, number = {4}, pages = {987--1012},
  doi = {10.1093/jrsssb/qkae022}
}

@article{feng2026optimal,
  author = {Feng, Qunqiang and Jin, Jiashun and Tian, Yaru and Yan, Ting},
  title = {Optimal estimators and tests for reciprocal effects},
  journal = {J. Am. Statist. Assoc.},
  year = {2026},
  doi = {10.1080/01621459.2026.2736314},
  note = {Advance online publication, doi:10.1080/01621459.2026.2736314}
}

@misc{feng2025regression,
  author = {Feng, Rui and Leng, Chenlei},
  title = {Regression analysis of reciprocity in directed networks},
  year = {2025},
  howpublished = {arXiv:2507.21469},
  doi = {10.48550/arXiv.2507.21469}
}

@article{shao2025reduction,
  author = {Shao, Meijia and Xia, Dong and Zhang, Yuan},
  title = {{$U$}-statistic reduction: Higher-order accurate risk control and statistical-computational trade-off},
  journal = {J. Am. Statist. Assoc.},
  year = {2025}, volume = {120}, number = {551}, pages = {1593--1606},
  doi = {10.1080/01621459.2024.2448029}
}

@misc{ju2026automated,
  author = {Ju, Haocheng and Gao, Guoxiong and Jiang, Jiedong and Wu, Bin and Sun, Zeming and Liu, Shurui and Chen, Leheng and Wang, Yutong and Wang, Yuefeng and Wang, Zichen and He, Wanyi and Wu, Peihao and Xiao, Liang and Liu, Ruochuan and Dai, Bryan and Dong, Bin},
  title = {Automated conjecture resolution with formal verification},
  year = {2026},
  howpublished = {arXiv:2604.03789},
  doi = {10.48550/arXiv.2604.03789},
  url = {https://arxiv.org/abs/2604.03789}
}

@article{holland1981exponential,
  author = {Holland, Paul W. and Leinhardt, Samuel},
  title = {An exponential family of probability distributions for directed graphs},
  journal = {J. Am. Statist. Assoc.},
  year = {1981}, volume = {76}, number = {373}, pages = {33--50},
  doi = {10.1080/01621459.1981.10477598}
}
\end{document}